\documentclass{article}
\usepackage{dcase2020,amsmath,graphicx,url,times,booktabs, tabularx}
\usepackage{amsfonts,subfigure,url}
\usepackage{algorithm}
\usepackage{multirow}
\usepackage{algpseudocode}
\usepackage{footmisc} 
\usepackage{color}

\title{Acoustic Scene Classification with Spectrogram Processing Strategies}

\name{Helin Wang$^{1}$, Yuexian Zou$^{1,2,*}$, Dading Chong$^{1}$ \thanks{* Yuexian Zou is the corresponding author.}}
\address{
  $^1$ADSPLAB, School of ECE, Peking University, Shenzhen, China\\
  $^2$Peng Cheng Laboratory, Shenzhen, China\\ 
  \{wanghl15,zouyx,1601213984\}@pku.edu.cn}

\begin{document}

\ninept
\maketitle

\begin{sloppy}

\begin{abstract}
Recently, convolutional neural networks (CNN) have achieved the state-of-the-art performance in acoustic scene classification (ASC) task.
The audio data is often transformed into two-dimensional spectrogram representations, which are then fed to the neural networks.
In this paper, we study the problem of efficiently taking advantage of different spectrogram representations through discriminative processing strategies.
There are two main contributions.
The first contribution is exploring the impact of the combination of multiple spectrogram representations at different stages, which provides a meaningful reference for the effective spectrogram fusion.
The second contribution is that the processing strategies in multiple frequency bands and multiple temporal frames are proposed to make fully use of a single spectrogram representation.
The proposed spectrogram processing strategies can be easily transferred to any network structures.
The experiments are carried out on the DCASE 2020 Task1 datasets,
and the results show that our method could achieve the accuracy of $81.8\%$ (official baseline: $54.1\%$) and $92.1\%$ (official baseline: $87.3\%$) on the officially provided fold 1 evaluation dataset of Task1A and Task1B, respectively.
\end{abstract}


\section{Introduction}
\label{sec:intro}
Acoustic scene classification (ASC) aims to classify audio as one of a set of categories such as home, street, and office \cite{virtanen2017computational},
and has become an important research in the fields of acoustic signal processing.
Detection and Classification of Acoustic Scenes and Events (DCASE) challenges organized by IEEE Audio and Signal Processing (AASP) Technical Committee are one of the biggest competitions for ASC task \cite{dcase2020web},
which attract an increasing number of participants each year.
ASC is still a challenging task because of continuous, periodic or aperiodic acoustic signals that interfere with the understanding of the scene \cite{pham2020robust}.

Recently, deep learning has accomplished many achievements in audio, image, and natural language processing.
The algorithms based on the convolutional neural networks (CNN) are dominant in ASC tasks \cite{barchiesi2015acoustic,han2017convolutional,salamon2017deep}, which show powerful ability to extract the robust time-frequency information from audio.
In order to obtain more discriminative information, different features for input were exploited.
Many researchers transform the audio data into two-dimensional spectrogram representations to be fed into the back-end networks, 
and different spectrogram representations have been studied, such as Mel Frequency Cepstral Coefficients (MFCC) \cite{mcloughlin2016speech}, constant-Q transform (CQT) \cite{zeinali2018convolutional}, Gammatone spectrogram (Gamma) \cite{ellis2009gammatone} and log-Mel spectrogram (Log-Mel) \cite{mcloughlin2016speech}.
Among them, Sakashita \textit{et al.} \cite{sakashita2018acoustic} proposed an ensemble of spectrograms based on adaptive temporal divisions,
and Seo \textit{et al.} \cite{seo2019acoustic} proposed ensemble systems of CNN with various pre-processed features. 
Pham \textit{et al.} \cite{pham2020robust} presented an encoder-decoder architecture to parallelly map three spectrograms and combine them in the middle layers of the networks.
In addition, Ngo \textit{et al.} \cite{ngo2020sound} provided a comprehensive analysis on five common types of spectrograms, and made spectrogram fusion in the first layer of the networks.
Several other works \cite{phaye2019subspectralnet,qiao2019sub} focused on how to better use a single representation (\textit{e.g.} Log-Mel).
With the inspiration that different frequency bands in a spectrogram contain distinct features, Phaye \textit{et al.} \cite{phaye2019subspectralnet} proposed a SubSpectralNet network which was able to extract discriminative information from sub-spectrograms.
Qiao \textit{et al.} \cite{qiao2019sub} truncated the whole spectrogram into different sub-spectrograms, and adopted a score level based fusion mechanism to jointly improve the classification accuracy.

In this paper, we study the effective methods to take advantage of the spectrogram representations.
Based on the convolutional neural networks (CNN), we propose several spectrogram processing strategies to obtain more discriminative information for ASC,
including the spectrogram processing strategie in multiple representations (SPSMR), multiple frequency bands (SPSMF), and multiple temporal frames (SPSMT).
More specifically, log-Mel spectrogram (Log-Mel), constant-Q transform (CQT), Gammatone spectrograms (Gamma) and Mel Frequency Cepstral Coefficients (MFCC) are used as the input to the networks.
Instead of the feature-level fusion for the spectrogram representations, four independent networks with different representations are applied before the decision-level fusion, 
which is the SPSMR.
In addition, we exploit to make decision on the subparts of a spectrogram rather than the whole spectrogram to improve the robustness,
\textit{i.e.} the spectrogram processing strategy in multiple frequency bands (SPSMF) and the spectrogram processing strategy multiple temporal frames (SPSMT).
Under the official fold 1 evaluation setup of DCASE 2020 Task1 \cite{Heittola2020}, our system could achieve $81.8\%$ accuracy with $0.694$ log loss in the Task1A evaluation set,
and $92.1\%$ accuracy with $0.312$ log loss in the Task1B evaluation set.

The remainder of this paper is organized as follows. 
Section 2 gives full particulars of a series of processing strategies.
Section 3 details the architectures of our networks. 
Section 4 presents the details of experiments and results, and
Section 5 concludes this paper.

\section{Spectrogram Processing Strategies}
\label{sec:method}

In this section, the conventional CNN-based method and our proposed spectrogram processing strategies are introduced, 
which are the spectrogram processing strategy in multiple representations (SPSMR),
the spectrogram processing strategy in multiple frequency bands (SPSMF),
and the spectrogram processing strategy in multiple temporal frames (SPSMT).
In addition, several other spectrogram fusion methods are also introduced.

\subsection{The Conventional CNN-based Method}
CNN-based methods were widely used in ASC task, and provided the state-of-the-art performance \cite{chen2019integrating,koutini2019cp}.
To be specific, given an audio clip, the two-dimensional time-frequency representation (\textit{e.g.} Log-Mel) is first extracted. 
Convolutional layers are then applied to the time-frequency representation $\boldsymbol{M} \in \mathbb{R}^{T \times F}$ to obtain the deep representation $\boldsymbol{M}^{'} \in \mathbb{R}^{c \times t \times f}$,
where $c$ denotes the number of the output channels.
\begin{equation}
  \boldsymbol{M}^{'}=f_{\mathrm{cnn}}\left(\boldsymbol{M} ; \theta_{\mathrm{cnn}}\right)
  \label{eq1}
\end{equation}
Here, $f_{\mathrm{cnn}}$ denotes the operation of the convolutional layers and  $\theta_{\mathrm{cnn}}$ denotes the model parameters of the convolutional layers.
The global pooling layer (\textit{e.g.} global average pooling) and fully-connected layers are then applied to obtain the predicted score of the classification.
Let $f_{\mathrm{gp}}$, $f_{\mathrm{fc}}$ be the operations of the global pooling layer and the fully-connected layers, respectively.
The predicted score $\hat{\boldsymbol{y}} \in \mathbb{R}^{N}$ (where $N$ denotes the number of categories) can be obtained by
\begin{equation}
  \hat{\boldsymbol{y}}=f_{\mathrm{fc}}\left(f_{\mathrm{gp}}\left(\boldsymbol{M}^{'}\right) ; \theta_{\mathrm{fc}}\right)
  \label{eq2}
\end{equation}
where $\theta_{\mathrm{fc}}$ denotes the model parameters of the fully-connected layers.

\begin{figure}[t] 
  \centering
  \subfigure[Early fusion (EF)]{\label{h1}
  \begin{minipage}[c]{0.5\textwidth}
    \vspace{0.3cm}
  \centering
  \includegraphics[width=\columnwidth]{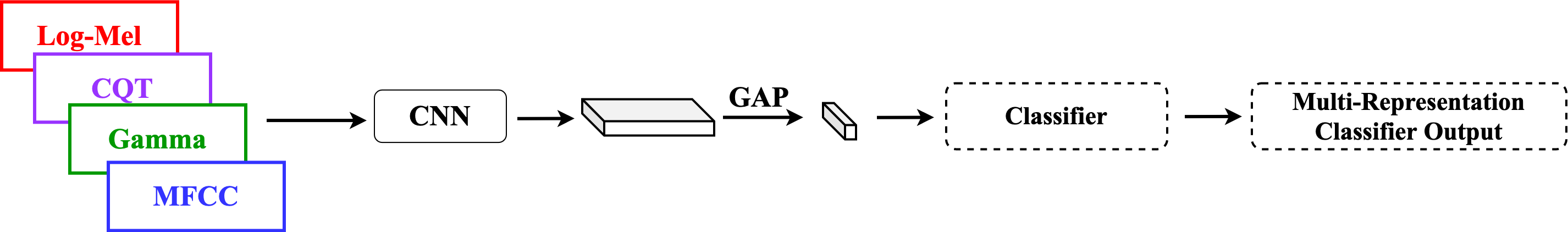}
  \end{minipage}
  }
  \subfigure[Middle fusion (MF)]{\label{h2}
  \begin{minipage}[c]{0.5\textwidth}
    \vspace{0.3cm}
  \centering
  \includegraphics[width=\columnwidth]{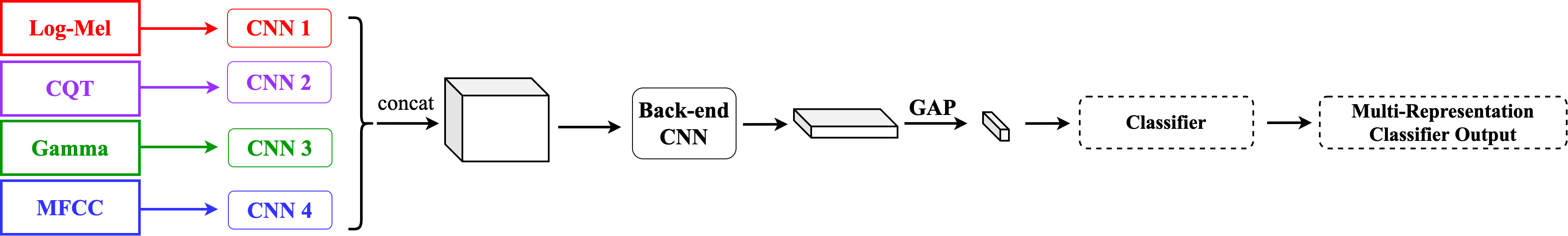}
  \end{minipage}
  }
  \subfigure[Late fusion (LF)]{\label{h3}
  \begin{minipage}[c]{0.5\textwidth}
    \vspace{0.3cm}
  \centering
  \includegraphics[width=\columnwidth]{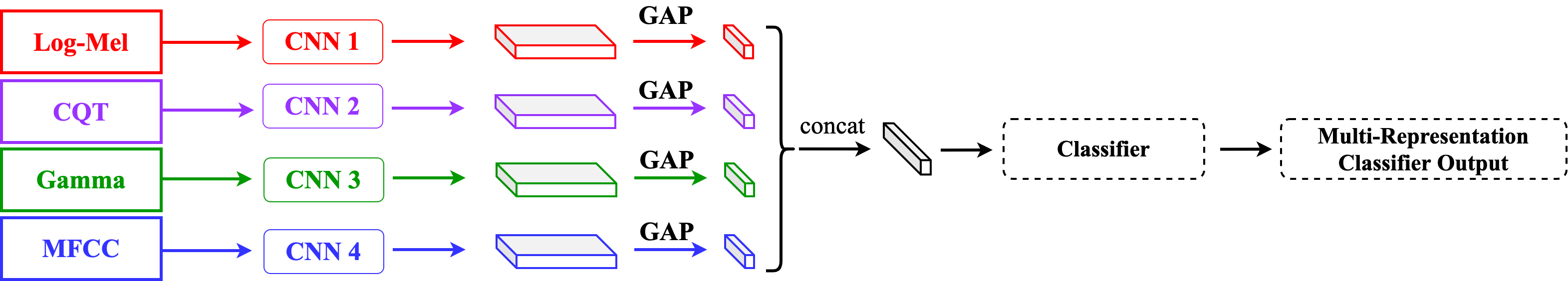}
  \end{minipage}
  }
  \subfigure[SPSMR]{\label{h4}
  \begin{minipage}[c]{0.5\textwidth}
    \vspace{0.3cm}
  \centering
  \includegraphics[width=\columnwidth]{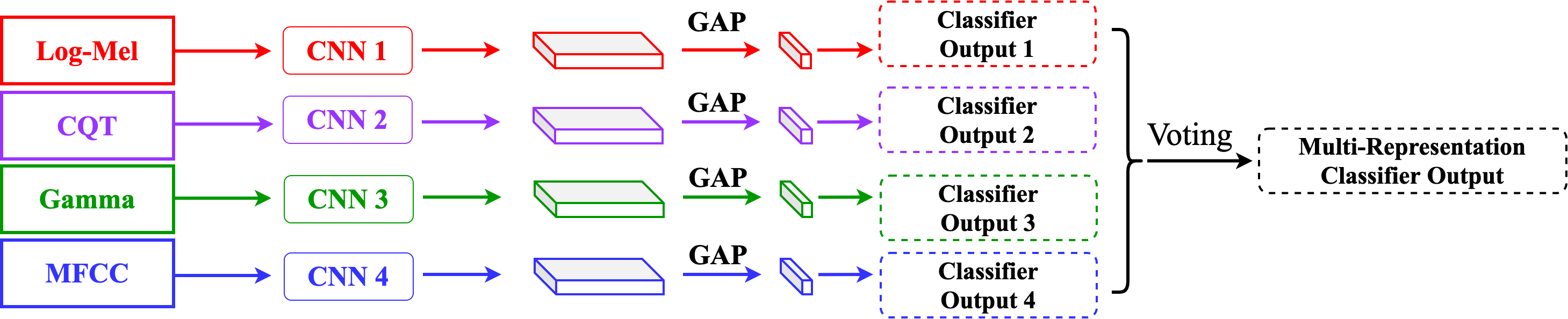}
  \end{minipage}
  }
  \caption{The illustration of different spectrogram fusion methods and our proposed SPSMR.}
  \label{fig:multiR}
\end{figure}

\subsection{The Spectrogram Processing Strategy in Multiple Representations}
\label{SPSMR}
Instead of inputting single representation to the networks, 
multiple representations can obtain more robust information from the raw audio.
Four types of spectrogram representations (\textit{i.e.} Log-Mel, CQT, Gamma, and MFCC) are used in our experiments.
As shown in Figure~\ref{h1}, one intuitive approach \cite{ngo2020sound} to apply multiple representations is inputting the multi-channel feature maps $\boldsymbol{M}^{*} \in \mathbb{R}^{n \times T \times F}$, where $n$ denotes the number of representations.
However, different representations have different characteristics and a single CNN network cannot model the differences effectively.
Other fusion methods \cite{pham2020robust,kong2019panns} could be the middle fusion in Figure~\ref{h2} or the late fusion in Figure~\ref{h3}, which jointly uses the multiple representations after several CNN layers or the whole CNN layers.
However, the same regions in different feature maps reflect different frequency information, which may cause the mismatched problem. 
In addition, a single classifier trained on the fusion features cannot take advantage of multiple representations well.

In order to overcome the above problems and make better use of the representations, 
we propose a novel spectrogram processing strategy in multiple representations (SPSMR),
which trains several independent CNN models based on different representations and then combines them by the decision-level fusion.
As shown in Figure~\ref{h4}, the predicted scores of the four models are averaged to obtain the final predicted score, which is also known as the average voting strategy.
\begin{equation}
  \hat{\boldsymbol{y}}=\frac{1}{4}\left(\hat{\boldsymbol{y}}_{1} + \hat{\boldsymbol{y}}_{2} + \hat{\boldsymbol{y}}_{3} + \hat{\boldsymbol{y}}_{4}\right)
  \label{eq3}
\end{equation}
where $\hat{\boldsymbol{y}}_{1}$, $\hat{\boldsymbol{y}}_{2}$, $\hat{\boldsymbol{y}}_{3}$, $\hat{\boldsymbol{y}}_{4}$
denotes the predicted score of CNN model with the input representation of Log-Mel, CQT, Gamma, and MFCC, respectively.
In this case, each independent model can focus on different representations and be more discriminative by the decision-level fusion.


\begin{figure}[t]
  \centering
  \centerline{\includegraphics[width=\columnwidth]{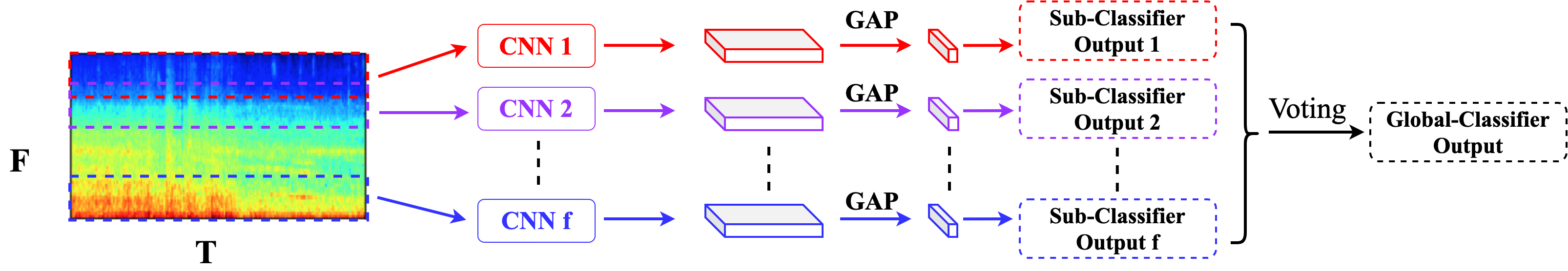}}
  \caption{The illustration of our proposed SPSMF.}
  \label{fig:multiF}
\end{figure}
\vspace{1.5cm}
\begin{figure}[t]
  \centering
  \centerline{\includegraphics[width=\columnwidth]{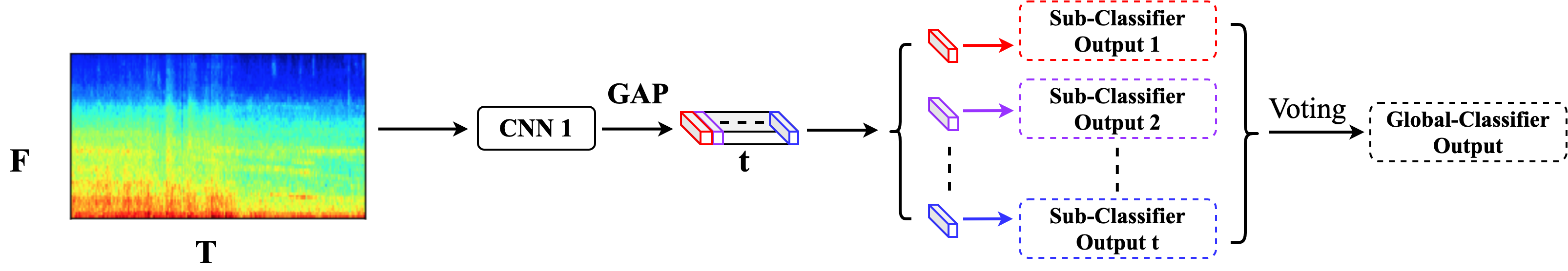}}
  \caption{The illustration of our proposed SPSMT.}
  \label{fig:multiT}
\end{figure}

\subsection{The Spectrogram Processing Strategy in Multiple Frequency Bands}
The spatial regions of the feature maps are treated equally in the conventional CNN-based methods, 
however, different acoustic scenes show different activity on the frequency bands \cite{phaye2019subspectralnet}.
Therefore, we take the sub-spectrograms \cite{phaye2019subspectralnet} as input and train several classifiers.
Different from \cite{phaye2019subspectralnet}, the final decision is made by the average voting strategy rather than training a global classifier, which shows better performance in our experiments.
As shown in Figure~\ref{fig:multiF}, for $f$ sub-spectrograms, the final score is obtained by
\begin{equation}
  \hat{\boldsymbol{y}}=\frac{1}{f} \sum_{i=1}^{f} \hat{\boldsymbol{y}}_{i}
  \label{eq4}
\end{equation}

\subsection{The Spectrogram Processing Strategy in Multiple Temporal Frames}
Several temporal devisions have been studied in \cite{sakashita2018acoustic} for ASC task to promote the generalization and robustness, including non division, non-overlap division and overlap division.
Among them, overlap division shows the best performance.
However, such division method only focused on the information within the window size of the division (\textit{e.g.} $2$s).
In this paper, a spectrogram processing strategy in multiple temporal frames (SPSMT) is proposed, which feeds the whole audio clip to the network and makes decision on each temporal frame after CNN.
Thus, the decision made by each frame could take into account the information of neighboring frames.
As shown in Figure~\ref{fig:multiT}, for the final feature map $\boldsymbol{M}^{'} \in \mathbb{R}^{c \times t \times f}$, 
global average pooling is applied to the frequency bands and the classifier is then applied to each temporal frames.
\begin{equation}
  \hat{\boldsymbol{y}}=\frac{1}{t} \sum_{i=1}^{t} \hat{\boldsymbol{y}}_{i}
  \label{eq5}
\end{equation}
Note that all the temporal frames share the same classifier in SPSMT, so that there are no extra parameters needed.

\section{Network Architectures}
\label{sec:NA}
Our base network architectures are shown in Table~\ref{t1}.
For DCASE 2020 Task1A, the network is a VGG \cite{simonyan2014very} style network, similar to \cite{kong2019cross}.
Batch normalization (BN) \cite{ioffe2015batch} and Rectified Linear Units (ReLU) \cite{nair2010rectified} are used following the convolutional operations.
Global pooling is applied after the last convolutional layer to obtain fixed-length vectors,
which is operated by the global average pooling in the frequency axis and the global max pooling in the temporal axis \cite{kong2019cross}.
Two fully-connected layers followed by a softmax function are then applied to obtain the prediction for classification.
Dropout with a ratio of $0.5$ is applied between the fully-connected layers.
While for DCASE 2020 Task1B, a tiny CNN model is employed to achieve the low complexity,
and other setups are the same as Task1A.

\section{Experiments}
\label{sec:ExA}
\subsection{Datasets and Metrics}
Our proposed method is evaluated on two novel datasets for ASC, 
\textit{i.e.} TAU Urban Acoustic Scenes 2020 Mobile, Development dataset (DCASE2020 1A) \cite{Heittola2020}
and TAU Urban Acoustic Scenes 2020 3Class, Development dataset (DCASE2020 1B) \cite{Heittola2020}.
DCASE2020 1A contains totally $64$h data from $10$cities and $9$ devices.
The dataset is provided with a training/test split in which $70$\% of the data for each device is included for training, 
$30$\% for testing.
The task targets generalization properties of systems across a number of different devices.
DCASE2020 1B contains data with a single device from $10$ cities. 
The total amount of audio is 40 hours, and audio is provided in binaural, $48$kHz $24$-bit format.
The task targets low complexity solutions for the classification problem.

\begin{figure}[t]
  \centering
  \centerline{\includegraphics[width=\columnwidth]{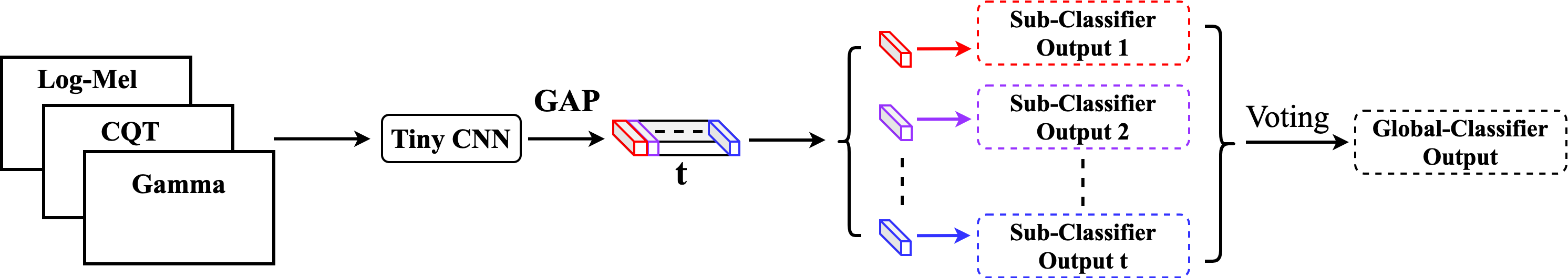}}
  \caption{The illustration of DCASE2020 1B network.}
  \label{fig:TB}
\end{figure}

\begin{table}[t]
  \begin{center}
  \caption{Network Architectures} \label{t1}
  \begin{tabular}{|c|c|}
      \hline
      \textbf{DCASE2020 1A} & \textbf{DCASE2020 1B}\\
      \hline
      Conv $3 \times 3 $ $@$ $64$, BN, ReLU & Conv $7 \times 7 $ $@$ $32$, BN, ReLU\\
      Conv $3 \times 3 $ $@$ $64$, BN, ReLU & \\
      \hline
      Avg Pooling $4 \times 2$ & Avg Pooling $4 \times 2$\\
      \hline
      Conv $3 \times 3 $ $@$ $128$, BN, ReLU & Conv $7 \times 7 $ $@$ $32$, BN, ReLU\\
      Conv $3 \times 3 $ $@$ $128$, BN, ReLU & \\
      \hline
      Avg Pooling $4 \times 2$ & Avg Pooling $4 \times 2$\\
      \hline
      Conv $3 \times 3 $ $@$ $256$, BN, ReLU & Conv $3 \times 3 $ $@$ $64$, BN, ReLU\\
      Conv $3 \times 3 $ $@$ $256$, BN, ReLU & \\
      \hline
      Avg Pooling $2 \times 2$ & Avg Pooling $2 \times 2$\\
      \hline
      Conv $3 \times 3 $ $@$ $512$, BN, ReLU & Conv $3 \times 3 $ $@$ $64$, BN, ReLU\\
      Conv $3 \times 3 $ $@$ $512$, BN, ReLU & \\
      \hline
      Global Pooling & Global Pooling\\
      \hline
      FC 512, ReLU & FC 200, ReLU\\
      \hline
      FC 10, softmax & FC 3, softmax\\
      \hline      
\end{tabular}
\label{t1}
\end{center}
\end{table}

Macro-average accuracy and multiclass cross-entropy (log loss) are used as our metrics.
In addition, we also report the model size.

\subsection{Experimental Setups}
For DCASE2020 1A, all the raw audios are resampled to $44.1$kHz
and fixed to the certain length of $10$s by zero-padding or truncating.
Log-Mel, CQT, Gamma, and MFCC are then extracted with window size 2048 ($46$ms) and hop length 512 ($11.6$ms).
The number of frequency bands are $40$, $64$, $64$ and $40$, respectively.
We test the single representation methods (\textit{i.e.} Log-Mel CNN, CQT CNN, Gamma CNN, and MFCC CNN),
and our proposed methods (\textit{i.e.} SPSMR, SPSMF, and SPSMT).

For DCASE2020 1B, in order to achieve the low complexity, early fusion (EF) is applied instead of SPSMR.
In addition, SPSMT is applied because of no extra parameters.
The overall architecture is shown in Figure~\ref{fig:TB}.
Three representations are used, \textit{i.e.} Log-Mel, CQT and Gamma,
and the number of frequency bands are all $64$.
Other experimental setups are the same as DCASE2020 1A.

In the training phase, the Adam algorithm \cite{kingma2014adam} is employed as the optimizer with the default parameters. 
The model is trained end-to-end with the initial learning rate of $0.001$ and the exponential decay rate of $0.91$ for each $200$ iterations.
Parameters of the networks are learned using the categorical cross entropy loss. 
Batch size is set to $64$ and training is terminated after $12000$ iterations. 
Data augmentation methods Mixup \cite{zhang2017mixup} is applied in our experiments to prevent the system from over-fitting and improve the performance.

\begin{table}[t]\footnotesize
  \begin{center}
  \caption{Comparison of accuracy and log loss on DCASE2020 1A dataset} \label{t2}
  \begin{tabular}{|c|c|c|c|}
     \hline
     \textbf{Model} & \textbf{Accuracy} & \textbf{Log loss} & \textbf{Model size} \\
     \hline
     DCASE2020 1A Baseline \cite{Heittola2020} & 54.1\% & 1.365 & 19.1 MB \\
     \hline
     Log-Mel CNN & 72.1\% & 0.879 & 18.9 MB \\
     CQT CNN & 71.2\% & 0.870 & 18.9 MB \\
     Gamma CNN & 76.1\% & 0.762 & 18.9 MB \\
     MFCC CNN & 63.6\% & 1.029 & 18.9 MB \\
     SPSMR & 79.4\% & 0.696 & 75.5 MB \\
     \hline
     Log-Mel CNN + SPSMF & 75.5\% & 1.135 & 94.4 MB\\
     CQT CNN + SPSMF & 74.5\% & 1.185 & 94.4 MB \\
     Gamma CNN + SPSMF & 78.8\% & 1.169 & 94.4 MB \\
     MFCC CNN + SPSMF & 60.9\% & 1.801 & 94.4 MB \\
     SPSMR + SPSMF & 80.9\% & 0.737 & 377.6 MB \\
     \hline
     Log-Mel CNN + SPSMT & 74.5\% & 0.987 & 18.9 MB \\
     CQT CNN + SPSMT & 73.3\% & 1.032 & 18.9 MB \\
     Gamma CNN + SPSMT & 78.2\% & 0.866 & 18.9 MB \\
     MFCC CNN + SPSMT & 67.6\% & 1.081 & 18.9 MB \\
     SPSMR + SPSMT & 79.7\% & 0.701 & 75.5 MB \\
     \hline
     \textbf{SPSMR + SPSMF + SPSMT} & \textbf{81.8\%} & \textbf{0.694} & 453.1 MB \\
     \hline

\end{tabular}
\end{center}
\end{table}

\begin{table}[t]\footnotesize
  \begin{center}
  \caption{Comparison of different spectrograms fusion methods on DCASE2020 1A dataset} \label{tx}
  \vspace{0.2cm}
  \begin{tabular}{|c|c|c|c|}
     \hline
     \textbf{Model} & \textbf{Accuracy} & \textbf{Log loss} & \textbf{Model size} \\
     \hline
     Log-Mel CNN & 72.1\% & 0.879 & 18.9 MB \\
     CQT CNN & 71.2\% & 0.870 & 18.9 MB \\
     Gamma CNN & 76.1\% & 0.762 & 18.9 MB \\
     MFCC CNN & 63.6\% & 1.029 & 18.9 MB \\
     \hline
     EF & 75.2\% & 0.852 & 18.9 MB \\
     MF & 77.0\% & 0.777 & 20.2 MB \\
     LF & 76.4\% & 0.698 & 72.5 MB \\
     \textbf{SPSMR} & \textbf{79.4\%} & \textbf{0.696} & 75.5 MB \\
     \hline

\end{tabular}
\end{center}
\end{table}

\subsection{Experimental Results}

Table~\ref{t2} demonstrates the experimental results of different models on DCASE2020 1A dataset.
Among them, SPSMR + SPSMF + SPSMT achieves the highest accuracy and the lowest log loss,
which shows that our proposed SPSMR, SPSMF, SPSMT can obviously improve the performance for ASC.
As for single representation methods (\textit{i.e.} Log-Mel CNN, CQT CNN, Gamma CNN and MFCC CNN),
Gamma CNN beats others.
SPSMR outperforms all the single representation methods, 
which evaluates the effectiveness of using multiple representations.
In addition, SPSMF and SPSMT can be directly applied to a single representation method or SPSMR,
and both show accuracy gains.
It is worth mentioning that SPSMT can improve the performance for ASC without extra parameters.

Furthermore, we compare different spectrogram fusion methods with our proposed SPSMR on DCASE2020 1A dataset,
and show the results in Table~\ref{tx}.
Apart from SPSMR, we report the results of four single representation methods (\textit{i.e.} Log-Mel CNN, CQT CNN, Gamma CNN and MFCC CNN) and three feature-level fusion methods (\textit{i.e.} EF, MF and LF),
which are introduced in Section~\ref{SPSMR}.
EF performs worse than the other fusion methods and even worse than a single representation method (\textit{i.e.} Gamma CNN).
This is because the less discriminative representation (\textit{e.g.} MFCC) for ASC may interfere other representations by EF.
In addition, MF performs better than LF, and becomes the best feature-level fusion method in our experiments.
It can be seen that our proposed SPSMR outperforms all the other methods, 
which shows the powerful ability of combining multiple representations.
We contribute this to the four independent networks in SPSMR rather than making fusion in the feature level, which allows each representation to be more discriminative.
The decision-level fusion by SPSMR also leads to the improvement of robustness
when a single representation shows poor discriminative for ASC.

Our proposed method is also evaluated on DCASE2020 1B dataset.
As presented in Table~\ref{t3}, EF+SPSMT outperforms the official baseline with the similar model size,
which shows that both EF and SPSMT are the powerful methods to improve the performance for ASC with few extra parameters. 
Different network structures are employed in DCASE2020 1A and 1B,
which shows that our proposed spectrogram processing strategies can fit different networks.

\begin{table}[t]\footnotesize
  \begin{center}
  \caption{Comparison of accuracy and log loss on DCASE2020 1B dataset} \label{t3}
  \begin{tabular}{|c|c|c|c|}
     \hline
     \textbf{Model} & \textbf{Accuracy} & \textbf{Log loss} & \textbf{Model size} \\
     \hline
     DCASE2020 1B Baseline \cite{Heittola2020} & 87.3\% & 0.437 & 450 KB \\
     \hline
     Log-Mel CNN & 87.5\% & 0.428 & 468 KB \\
     Log-Mel CNN + SPSMT & 90.8\% & 0.371 & 468 KB \\
     \textbf{EF + SPSMT} & \textbf{92.1\%} & \textbf{0.312} & 491 KB \\
     \hline
\end{tabular}
\end{center}
\end{table}

\section{Conclusion}
\label{sec:Con}
In this paper, 
three spectrogram processing strategies (\textit{i.e.} SPSMR, SPSMF, and SPSMT) have been proposed to make better use of the spectrogram representations and greatly improved the performance for ASC task.
These strategies were designed to fit the audio characteristics, 
and can be directly applied to other neural networks.
We believe that the proposed strategies can offer good generalization properties for other audio processing tasks.
The code for the experiments is available.\footnote{\url{https://github.com/WangHelin1997/DCASE-2020-Task1A-Code}}

\section{ACKNOWLEDGMENT}
\label{sec:ack}
This work was partially supported by Shenzhen Science \& Technology Fundamental Research Programs 
(No: JCYJ20170817160058246 \& JCYJ20180507182908274). 
Thanks for the code\footnote{\url{https://github.com/qiuqiangkong/dcase2019_task1}} provided by Qiuqiang Kong.

\bibliographystyle{IEEEtran}
\bibliography{refs}

%
%
%
%
%
%
%
%
%

\end{sloppy}
\end{document}